\documentclass{PoS}

\title{
\vspace*{-1.5cm}
\begin{flushright}
\texttt{\footnotesize CERN-PH-TH-2017-022}\\
\end{flushright}
\vspace*{0.5cm}
Continuum limit and universality \\ of the Columbia plot}

\ShortTitle{Universality and Columbia plot}

\author{\speaker{Philippe de Forcrand}  \\ 
  Institut f\"ur Theoretische Physik, ETH Z\"urich, CH-8093 Z\"urich, Switzerland \\
  CERN, Theory Division, CH-1211 Geneva 23, Switzerland \\
        E-mail: \email{forcrand@phys.ethz.ch}}

\author{Massimo D'Elia \\
  INFN - Sezione di Pisa, Largo Pontecorvo 3, I-56127 Pisa, Italy \\ 
  Dipartimento di Fisica dell'Universit\`a di Pisa, Largo Pontecorvo 3, I-56127 Pisa, Italy \\
        E-mail: \email{massimo.delia@unipi.it}}

\abstract{
Results on the thermal transition of QCD with 3 degenerate flavors, in the lower-left corner
of the Columbia plot, are puzzling.
The transition is expected to be first-order for massless quarks, and to remain so for
a range of quark masses until it turns second-order at a critical quark mass. 
But this critical quark mass and resulting ''pion'' mass disagree violently between Wilson 
and staggered fermions at finite lattice spacing, and decrease sharply with the lattice spacing, 
for staggered fermions at least.
To clarify this puzzle and eliminate potential systematic effects from rooting, we study
the 4-flavor theory with staggered fermions, on lattices with 4 to 10 time-slices.
Our results are qualitatively similar to the 3-flavor case, so that rooting is not
an issue. However, dramatic cutoff effects are visible, even on our finest lattices.
Universality implies that cutoff effects for Wilson fermions are even more dramatic.
In order to obtain a first-order thermal transition in the continuum theory, 
extremely light quarks are needed.
}

\FullConference{34th annual International Symposium on Lattice Field Theory\\
		24-30 July 2016\\
		University of Southampton, UK}

\begin{document}

\section{Introduction}

\begin{figure}
\centerline{
\hspace*{-3.0cm}
\includegraphics[width=0.45\textwidth]{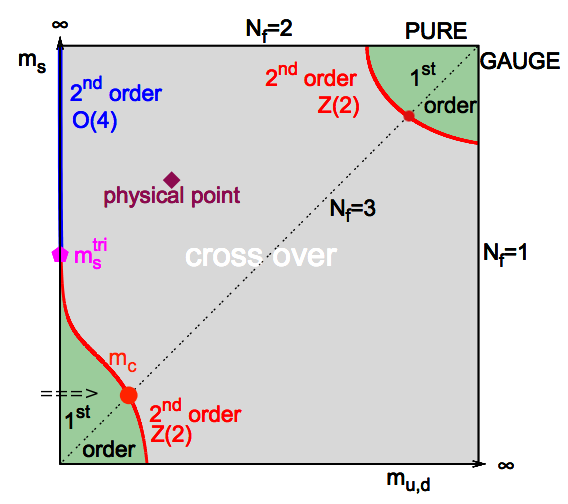}
\includegraphics[width=0.30\textwidth,viewport= 0 -20 200 200]{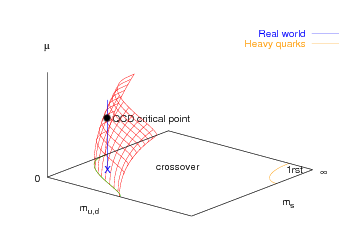}
}
\caption{({\em left}) Columbia plot: we focus on the $N_f=3$ chiral critical point (arrow),
and its $N_f=4$ analogue; ({\em right}): adding a vertical axis for the chemical potential $\mu$,
a possible QCD chiral critical point occurs when
the surface swept by the $\mu=0$ chiral critical line intersects the physical quark masses'
vertical line.}
\end{figure}

The QCD phase diagram summarizes the various behaviors of QCD as a function
of temperature $T$ and matter density, or equivalently quark chemical potential $\mu$.
Since the chiral and the center symmetry, which play crucial roles in the phase
diagram, are both explicitly broken in QCD by the quark masses, it is useful
to consider these masses as QCD parameters: $m_{u,d}$ for the two light quark masses
considered degenerate for simplicity, and $m_s$ for the strange quark mass.
Our expectations for the $\mu=0$ phase diagram, projected along the $T$-direction,
are contained in the ``Columbia plot'' Fig.~1 (left).

The upper-right and lower-left corners of the Columbia plot are simpler to analyze: \\
- In the first, all quarks are infinitely massive. They decouple, and the resulting
$SU(N_c=3)$ Yang-Mills theory obeys the global $Z(3)$ center symmetry, which is spontaneously broken
at high temperature via a first-order transition. \\
- In the second, all quarks are massless, and the theory obeys the global $SU(N_f=3)$ chiral symmetry,
which is spontaneously broken at low temperature and restored at high temperature. 
For 3 massless flavors or more, one expects symmetry
restoration to occur via a first-order transition~\cite{Pisarski_Wilczek}, because no
$3d$ $SU(N_f), N_f \geq 3$, second-order universality class is known~\cite{Vicari}. \\

In the middle of the Columbia plot, where both symmetries are badly broken explicitly,
Monte Carlo simulations indicate an analytic crossover as $T$ is raised.
Thus, there must exist two critical, second-order lines separating the two first-order
regions above from the central crossover region. Because no particular symmetry is at play
along these critical lines, their universality class should be that of a $3d$ $\phi^4$ theory,
i.e. that of the $3d$ Ising model. 

A simple way to pin down the location of these two critical lines is to consider the $N_f=3$
case, with all quark masses equal, shown as the diagonal of the Columbia plot.
Two critical quark masses should be observed, to be determined with high precision via
Monte Carlo simulations. In practice, it is difficult to adopt a reference scale,
since an $N_f=3$ theory is a distortion of real-world QCD. So, it is convenient to
trade the critical quark mass for the ratio of the corresponding $T=0$ ``pion'' mass $m_\pi^c$ over 
the transition temperature $T_c$: $m_\pi^c / T_c$ is of order 1 for real-world QCD,
and allows to separate the regime of "light" and that of "heavy" quarks.

Besides increasing our fundamental knowledge, the quantitative determination of the Columbia plot
is useful when considering the effect of a chemical potential $\mu$, which can be viewed
as an additional vertical axis. The two $\mu=0$ critical lines discussed above will sweep
critical surfaces as $\mu$ is turned on. The chiral critical surface, near the lower-left corner,
may bend away from the origin, and reach the physical quark mass values for a sufficiently
large $\mu$: this signals the presence of a QCD chiral critical point, as in Fig.~1 (right).
Or this critical surface may bend the other way, and there may be no chiral critical point~\cite{PdF_OP}.
Even in the first case, reaching the critical point will require more bending if the $\mu=0$
critical line corresponds to smaller quark masses. Thus, an accurate determination
of this critical line is an important ingredient to shape our understanding of the
finite-density properties of QCD.
This can all be studied at $\mu=0$, without having to face the ``sign problem'' present
at non-zero chemical potential.

The importance of pinning down the $\mu=0, N_f=3$ critical points has been recognized.
The technical difficulty is to control the approach to the continuum limit.
For the heavy quark case, the masses are $\gtrsim {\cal O}(1)$ GeV, which requires
fine lattices to avoid UV cutoff effects. Numerical work so far has focused on coarse
lattices with $N_t=4$ time-slices~\cite{heavy}. Such difficulties are absent in the light quark case, so that
one would expect reasonable accuracy for lattice spacings ${\cal O}(0.1)$ fm, i.e. $N_t \gtrsim 8$.
However, one observes $(i)$ large cutoff effects ($\sim 30\%$) for $N_t=4,6,8$~\cite{Ukawa1},
and $(ii)$ enormous discrepancies (a factor of $\sim$ four!) between staggered and Wilson fermions at these values of $N_t$.

\begin{table}
\begin{center}
\begin{tabular}{ccccc}
Action & $N_t$ & $m_\pi^c$ & Ref. & Year \\ 
\hline
standard staggered & 4 & $\sim 290$ MeV & \cite{KS_Nt4} & 2001 \\ 
p4 staggered & 4 & $\sim  67$ MeV & \cite{p4_Nt4} & 2004 \\ 
standard staggered & 6 & $\sim 150$ MeV & \cite{KS_Nt6} & 2007 \\ 
HISQ staggered & 6 & $\lesssim 45$ MeV & \cite{HISQ} & 2011 \\ 
stout staggered & 4-6 & could be zero & \cite{Varnhorst} & 2014 \\ 
\hline
Wilson-clover & 6-8 & $\sim 300$ MeV & \cite{Ukawa1} & 2014 \\ 
Wilson-clover & 4-10 & $\sim 100$ MeV & \cite{Ukawa2} & 2016 \\
\hline
\end{tabular}
\end{center}
\caption{Summary of previous studies of the $N_f=3$, $\mu=0$ chiral critical point -- adapted from \cite{Varnhorst}.
The general trend is: finer lattices and/or improved actions drive the critical ''pion'' mass down;
Wilson fermions favor much heavier values. The last line (Wilson, $N_t=10$) was presented at the Lattice
conference~\cite{Ukawa2}. Other, related studies have kept fixed to their physical value 
the strange quark mass~\cite{ms_fixed}, or the ratio $m_s/m_{u,d}$~\cite{Endrodi}.}
\end{table}

Table I, adapted from \cite{Varnhorst}, summarizes the results of past studies, and adds some results
presented at the Lattice conference~\cite{Ukawa2}. The general trend is: the more one approaches 
the continuum limit, by decreasing the lattice spacing or by improving the action, the softer the
transition, and the lighter
the critical ``pion'' is. The most remarkable case is that of \cite{Varnhorst}: using a stout-improved
action with staggered fermions on an $N_t=4$ lattice, no sign of a first-order transition was
found, down to arbitrarily small quark masses! In contrast, with Wilson fermions Ref.~\cite{Ukawa1}
finds a critical ``pion'' mass of about 300 MeV after extrapolating from $N_t=6$ and $8$.

Thus, one is led to mistrusting the staggered simulations for two reasons: they indicate 
a very small critical ``pion'' mass, consistent with zero; and they disagree strongly with
Wilson fermion results, both at finite lattice spacing and after continuum extrapolation.

One plausible culprit for these two puzzles might be rooting. With staggered fermions,
the Dirac determinant is raised to the power $3/4$ to mimic $N_f=3$ degenerate flavors.
The danger of this procedure has been pointed out~\cite{rooting}: the consensus is that
danger appears when the chiral limit is approached first, before the continuum limit
is taken. This is potentially the case here, since the quark masses needed for criticality
quickly approach zero as $N_t$ is increased.

To eliminate a possible issue with rooting, we have studied the case of $N_f=4$ degenerate
flavors: no rooting is required, and the thermal transition in the chiral limit is expected
to be first-order as for $N_f=3$. Actually, a naive counting of the degrees of freedom
suggests that the first-order transition
will be stronger for $N_f=4$ than for $N_f=3$, so that the critical ``pion'' mass will be
heavier, thus reducing the computing cost of the simulations.

\begin{figure}[t]
\centerline{
\includegraphics[width=0.75\textwidth]{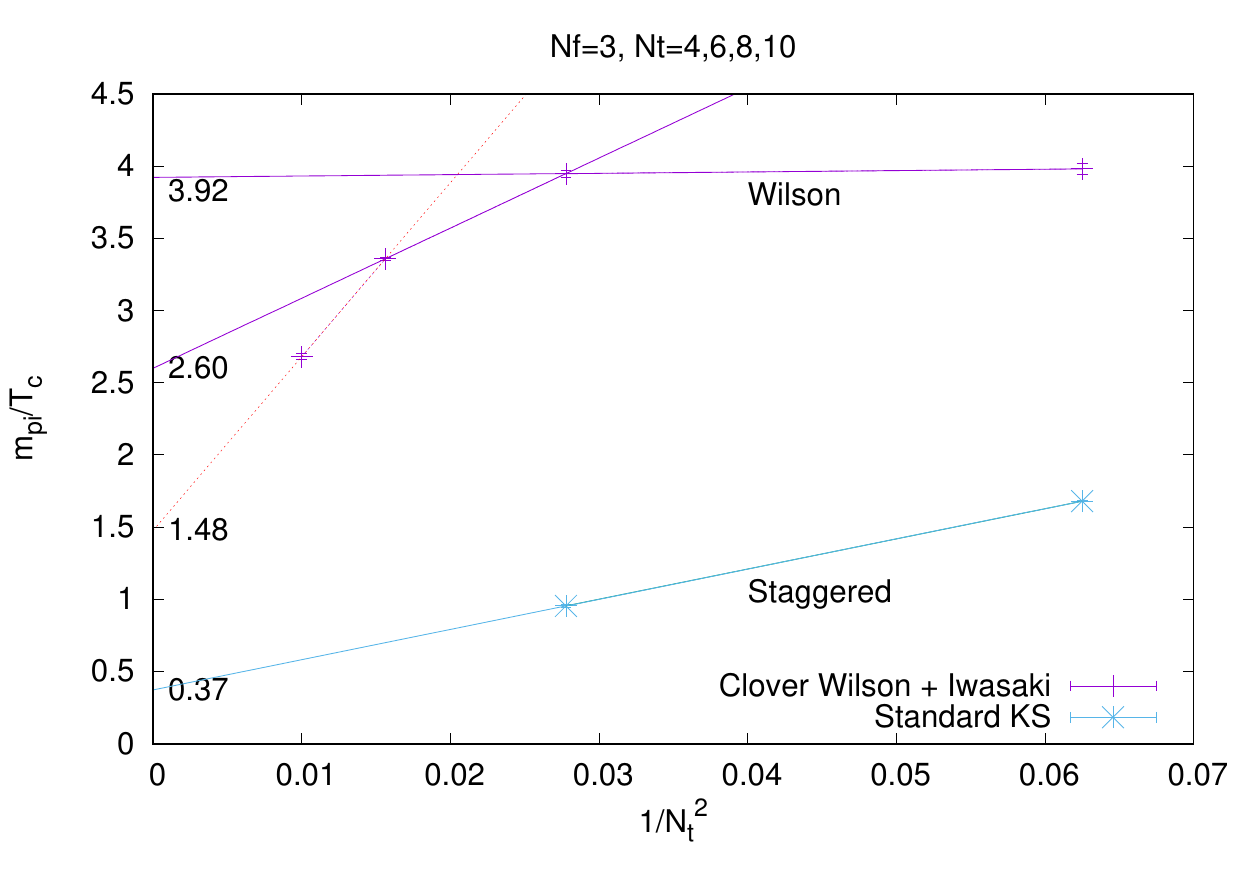}
}
\caption{$N_f=3$ comparison of $m_\pi^c/T_c$ as a function of the lattice spacing $a^2$,
between standard staggered fermions~\cite{KS_Nt6} and non-perturbatively
improved Wilson fermions~\cite{Ukawa1}. 
The $N_t=4$ and $6$ Wilson fermion data would mistakenly suggest small lattice artifacts.
The $N_t=10$ Wilson point was presented at the Lattice 
conference~\cite{Ukawa2}. The numbers along the $y$-axis indicate the results of linear
extrapolations in $a^2$.}
\end{figure}

\section{Results: $N_f=4$}

As argued above, we have simulated standard staggered fermions with $N_f=4$ flavors
and Wilson plaquette action, in order to bypass potential harmful effects of rooting and
to keep computing costs down.
The numerical simulations have been 
performed using a code running on GPUs~\cite{gpupaper}.

For successive values of $N_t=4, 6, 8$ and $10$, we have simulated lattices of spatial size 
$N_s \geq 2 N_t$, and determined the light bare quark mass $m_q^c$ for which the finite-temperature
chiral transition is second-order. Following \cite{PdF_OP}, the order of the phase transition was
established by monitoring the Binder cumulant $B_4(\bar\psi \psi)$, where 
$B_4(X) \equiv \frac{\langle (\delta X)^4 \rangle}{\langle (\delta X)^2 \rangle^2}$ and
$\delta X \equiv X - \langle X \rangle$. Near criticality, $B_4$ should be a function of
the ratio $L_s/\xi$ of the spatial lattice size over the spatial correlation length, which
diverges as $|m_q - m_q^c|^{-\nu}$. The critical value 1.604.. and the critical exponent $\nu \approx 0.63$
are known from the $3d$ Ising universality class. Thus, one can expand $B_4(m_q)$ near $m_q^c$ as
\begin{equation}
B_4(m_q) = 1.604 + c_1 (m_q - m_q^c) N_s^{1/\nu} + {\cal O}((m_q - m_q^c)^3)
\end{equation}
An illustrative fit (including cubic terms) is shown in Fig.~3 (left) for $N_t=4$.

For a given value of $N_t$, this procedure determines the bare parameters $a m_q^c$ and $\beta$ 
required for criticality. A zero-temperature simulation is then performed, at these parameters,
to determine the $T=0$ pion mass $(a m_\pi^c)(N_t)$. Finally, one obtains the physically meaningful
ratio $m_\pi^c/T_c = N_t a m_\pi^c$, and repeats this procedure for successive values of $N_t$. 
These successive ratios are shown in Fig.~3 (right).

The zero-temperature simulations require large lattices $(N_s^0)^3 \times N_t^0$, both spatially 
(to maintain $(m_\pi L_s) \gg 1$) and temporally (to achieve $T\approx 0$). 
To reduce the computing effort, we chose $N_s^0 \geq 2 N_t$ and $N_t^0 \geq 4 N_t$, i.e.
we used a $20^3 \times 40$ ``zero-temperature'' lattice in combination with a $20^3 \times 10$ 
finite-temperature lattice. We are aware that our choice is only marginally satisfactory, and
causes systematic errors in the extracted pion mass. However, Fig.~3 (right) shows variations
of order 100\% as $N_t$ is varied, which makes our systematic errors negligible in comparison.

\begin{figure}
\centerline{
\includegraphics[width=0.55\textwidth]{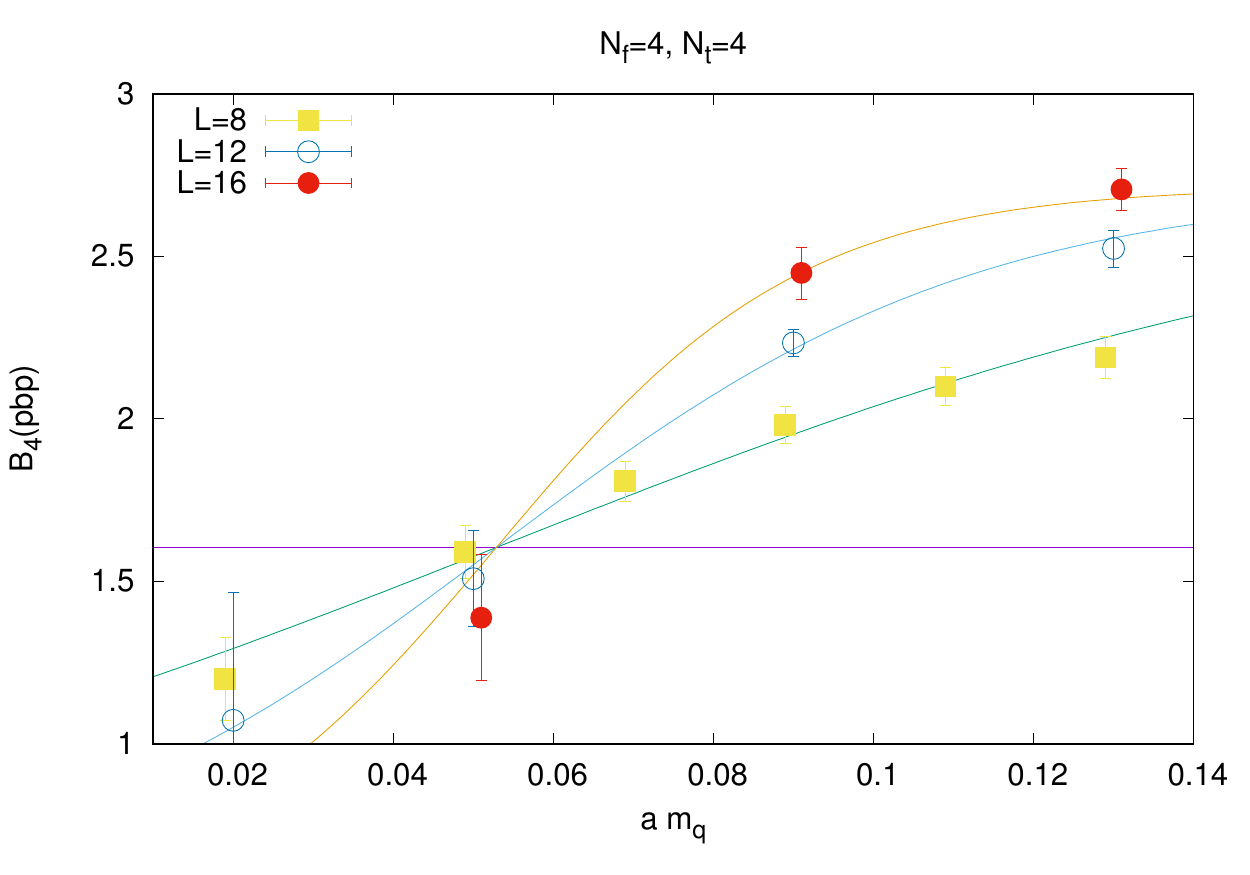}
\includegraphics[width=0.55\textwidth]{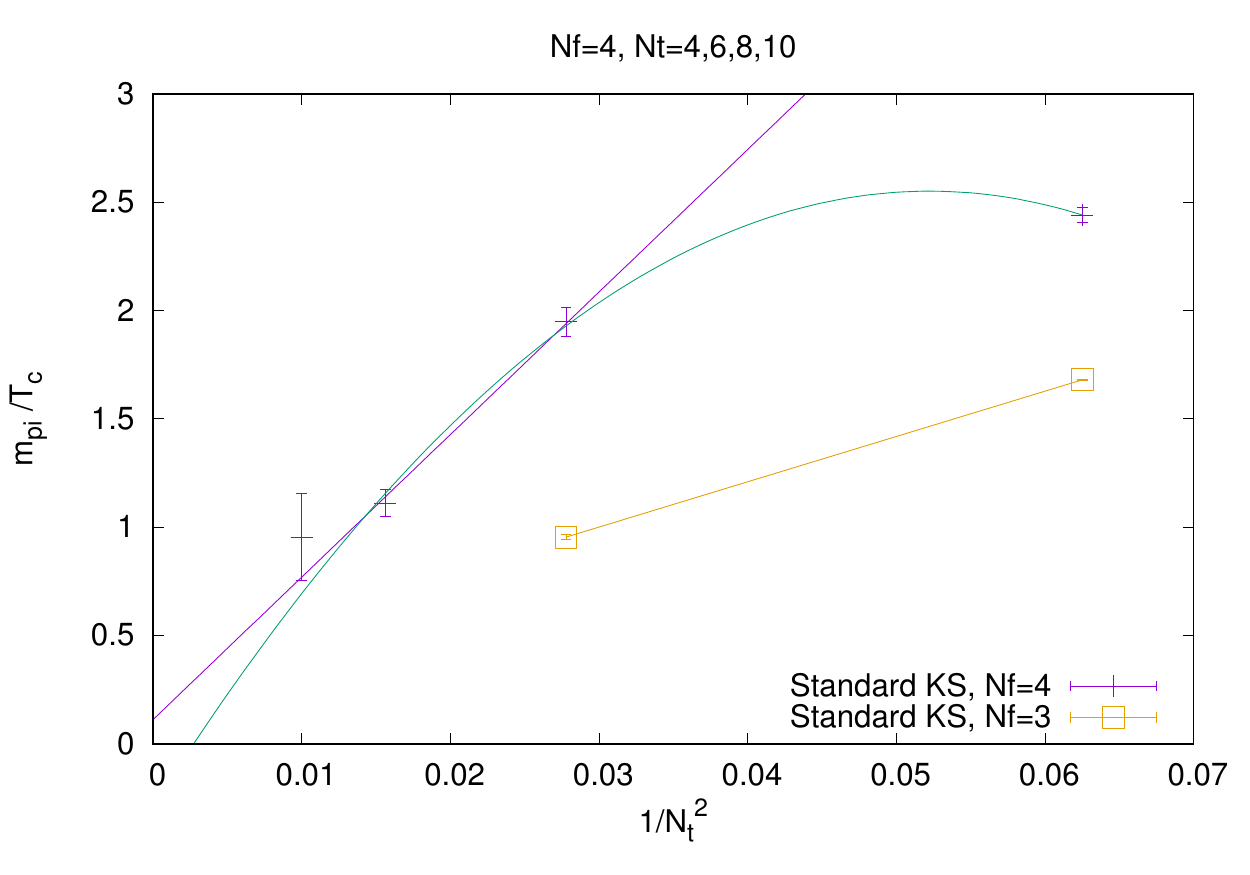}
}
\caption{({\em left}): finite-size scaling of the Binder cumulant $B_4(\bar\psi \psi)$
with the bare quark mass, here for $N_t=4$.
({\em right}): $N_f=4$ variation of $m_\pi^c/T_c$ as a function of the lattice spacing $a^2$, with 
linear and quadratic extrapolations. $N_f=3$ staggered results are shown for comparison: the critical
``pion'' mass is smaller, as expected.}
\end{figure}

\section{Discussion}

Our $N_f=4$ results Fig.~3 (right) are qualitatively similar to previous $N_f=3$ results:
the dramatic reduction or $m_\pi^c/T_c$ as the lattice spacing is reduced is still present, 
and thus not related to rooting.
Actually, a similar reduction is now visible in the Wilson data as well, with the new $N_t=10$ 
point~\cite{Ukawa2} (see Fig.~2).
Therefore, there is no reason to doubt the universality of the continuum limit:
the Wilson and the staggered values should converge as $N_t$ is increased. 

What is remarkable, however, is how slow this convergence is, even with a non-perturbatively improved
action as in the Wilson case!
Note that $N_t=10$, from which the continuum value of $m_\pi^c/T_c$ will probably differ by a factor 2
or more, corresponds to a lattice spacing $a \sim 0.13$ fm. State-of-the-art thermodynamic studies
use a maximum number of $N_t=16$ time-slices.

Let us speculate on the reason for such large cutoff effects. Taste-breaking could be the explanation:
$N_f=4$ staggered fermions possess 16 ''pions'', but only one of them is really light, and the 15 others
become degenerate with it in the continuum limit only. Thus, as the lattice spacing is reduced, the
number of light pions effectively increases. But this should make the transition stronger, not weaker as observed.
Moreover, the opposite occurs with Wilson fermions: there, the doublers become heavier toward
the continuum limit. Nevertheless, Wilson and staggered fermions both lead to the transition
becoming softer in the continuum limit. Thus, the explanation must reside elsewhere.

Perhaps cutting off all but the $N_t$ lowest Matsubara frequencies has larger than expected consequences.
One simple exercise consists of calculating the pressure of a free massless boson on the lattice, and
comparing it with the continuum Stefan-Boltzmann law~\cite{Karsch_SB}. An instructive figure can
be found in Fig.~2 of \cite{Helvio}. It shows that the lattice pressure can easily differ from the 
continuum one by factors well beyond 10. The pressure deficit due to the Matsubara cutoff is less 
pronounced if the boson is more massive, so that the cutoff will extend the parameter regime 
of the confined phase, and push the critical ''pion'' mass upward.

If this guess is correct, then it is essential to reduce the {\em temporal} lattice spacing,
not so much the spatial one. Lattice actions with anisotropic couplings would afford an economical
approach to the continuum limit. For the measurement of $m_\pi^c/T_c$, the non-perturbative tuning 
of the (gauge and fermion) anisotropy coefficients is not needed, as long as the continuum limit is consistent.
Alternatively, one could improve the action so as to approach the Stefan-Boltzmann law better,
in the spirit of the p4-improved action~\cite{p4}.

Controlling the continuum extrapolation of the $N_f=4$ finite temperature transition will
serve us for the $N_f=2+1$ case as well. There, unexpected results have been obtained for the
upper-left corner of the Columbia plot: the thermal transition appears to be first-order -- on 
a coarse, $N_t=4$ lattice~\cite{Nf2}. Lattice corrections should also be carefully considered
in the search for a conformal window in the number of flavors: it has been proposed to identify
the lower edge of this window as the number of massless flavors for which the critical temperature
of the chiral phase transition reaches zero~\cite{Pallante} -- there too, lattice corrections
may well be very significant.

Finally, it is clear that the continuum value of $m_\pi^c/T_c$, for $N_f=4$ and even more so for $N_f=3$,
is going to be extremely small. At present, continuum extrapolations as in Fig.~3 (right) are
compatible with a zero value. Could it actually be exactly zero, in contradiction with the
predictions of \cite{Pisarski_Wilczek} ?

\acknowledgments

Numerical simulations have been performed 
on a GPU farm located at the INFN Computer Center in Pisa and on the
QUONG cluster in Rome.


\begin{thebibliography}{99}


\bibitem{Pisarski_Wilczek}
  R.~D.~Pisarski and F.~Wilczek,
  Phys.\ Rev.\ D {\bf 29} (1984) 338.

\bibitem{Vicari}
  A.~Butti, A.~Pelissetto and E.~Vicari,
  JHEP {\bf 0308} (2003) 029
  doi:10.1088/1126-6708/2003/08/029
  [hep-ph/0307036].

\bibitem{PdF_OP}
  P.~de Forcrand and O.~Philipsen,
  Nucl.\ Phys.\ B {\bf 673} (2003) 170
  [hep-lat/0307020];
  JHEP {\bf 0811} (2008) 012
  [arXiv:0808.1096 [hep-lat]].

\bibitem{heavy}
  C.~Alexandrou, A.~Borici, A.~Feo, P.~de Forcrand, A.~Galli, F.~Jegerlehner and T.~Takaishi,
  Phys.\ Rev.\ D {\bf 60} (1999) 034504
  [hep-lat/9811028].
  H.~Saito {\it et al.} [WHOT-QCD Collaboration],
  Phys.\ Rev.\ D {\bf 84} (2011) 054502
   Erratum: [Phys.\ Rev.\ D {\bf 85} (2012) 079902]
  [arXiv:1106.0974 [hep-lat]].
  M.~Fromm, J.~Langelage, S.~Lottini and O.~Philipsen,
  JHEP {\bf 1201} (2012) 042
  [arXiv:1111.4953 [hep-lat]].

\bibitem{Ukawa1}
  X.~Y.~Jin, Y.~Kuramashi, Y.~Nakamura, S.~Takeda and A.~Ukawa,
  Phys.\ Rev.\ D {\bf 91} (2015) no.1,  014508
  [arXiv:1411.7461 [hep-lat]].

\bibitem{KS_Nt4}
  F.~Karsch, E.~Laermann and C.~Schmidt,
  Phys.\ Lett.\ B {\bf 520} (2001) 41
  [hep-lat/0107020].

\bibitem{p4_Nt4}
  F.~Karsch, C.~R.~Allton, S.~Ejiri, S.~J.~Hands, O.~Kaczmarek, E.~Laermann and C.~Schmidt,
  Nucl.\ Phys.\ Proc.\ Suppl.\  {\bf 129} (2004) 614
  [hep-lat/0309116].

\bibitem{KS_Nt6}
  P.~de Forcrand, S.~Kim and O.~Philipsen,
  PoS LAT {\bf 2007} (2007) 178
  [arXiv:0711.0262 [hep-lat]].

\bibitem{HISQ}
  H.-T.~Ding, A.~Bazavov, P.~Hegde, F.~Karsch, S.~Mukherjee and P.~Petreczky,
  PoS LATTICE {\bf 2011} (2011) 191
  [arXiv:1111.0185 [hep-lat]].
  A.~Bazavov, H.-T.~Ding, P.~Hegde, F.~Karsch, E.~Laermann, S.~Mukherjee, P.~Petreczky and C.~Schmidt,
  arXiv:1701.03548 [hep-lat].

\bibitem{Varnhorst}
  L.~Varnhorst,
  PoS LATTICE {\bf 2014} (2015) 193.

\bibitem{Ukawa2}
  S.~Takeda, X.~Y.~Jin, Y.~Kuramashi, Y.~Nakamura and A.~Ukawa,
  arXiv:1612.05371 [hep-lat].

\bibitem{ms_fixed}
  A.~Bazavov {\it et al.},
  Phys.\ Rev.\ D {\bf 85} (2012) 054503
  [arXiv:1111.1710 [hep-lat]].

\bibitem{Endrodi}
  G.~Endrodi, Z.~Fodor, S.~D.~Katz and K.~K.~Szabo,
  PoS LAT {\bf 2007} (2007) 182
  [arXiv:0710.0998 [hep-lat]].

\bibitem{rooting}
  M.~Creutz,
  PoS LAT {\bf 2007} (2007) 007
  [arXiv:0708.1295 [hep-lat]].
  C.~Bernard, M.~Golterman, Y.~Shamir and S.~R.~Sharpe,
  Phys.\ Rev.\ D {\bf 77} (2008) 114504
  [arXiv:0711.0696 [hep-lat]].

\bibitem{gpupaper} 
  C.~Bonati, G.~Cossu, M.~D'Elia and P.~Incardona,
  Comput.\ Phys.\ Commun.\  {\bf 183}, 853 (2012)
  [arXiv:1106.5673 [hep-lat]].

\bibitem{Karsch_SB}
  J.~Engels, F.~Karsch and H.~Satz,
  Nucl.\ Phys.\ B {\bf 205} (1982) 239.

\bibitem{Helvio}
P.~de Forcrand, P.~Romatschke, W.~Unger and H.~Vairinhos,
\bibitem{deForcrand:2017obe}
  P.~de Forcrand, P.~Romatschke, W.~Unger and H.~Vairinhos,
  PoS LATTICE {\bf 2016} (2017) 086,
  arXiv:1701.08324 [hep-lat].

\bibitem{p4}
  U.~M.~Heller, F.~Karsch and B.~Sturm,
  Phys.\ Rev.\ D {\bf 60} (1999) 114502
  [hep-lat/9901010].

\bibitem{Nf2}
  C.~Bonati, P.~de Forcrand, M.~D'Elia, O.~Philipsen and F.~Sanfilippo,
  Phys.\ Rev.\ D {\bf 90} (2014) no.7,  074030
  [arXiv:1408.5086 [hep-lat]].

\bibitem{Pallante}
  K.~Miura, M.~P.~Lombardo and E.~Pallante,
  Phys.\ Lett.\ B {\bf 710} (2012) 676
  [arXiv:1110.3152 [hep-lat]].

\end{thebibliography}
\end{document}